Distinguishing between paediatric brain tumour types using multi-parametric magnetic resonance imaging and machine learning: a multi-site study.

**Authors:** James T. Grist[1], Stephanie Withey[1,2,3], Lesley MacPherson[4], Adam Oates[4], Stephen Powell[1], Jan Novak[2,5], Laurence Abernethy[6], Barry Pizer[7], Richard Grundy[8], Simon Bailey[9], Dipayan Mitra[10], Theodoros N. Arvanitis[1,2,11], Dorothee P. Auer[12,13], Shivaram Avula[6], Andrew C Peet[1,2].

1. Institute of Cancer and Genomic Sciences, School of Medical and Dental Sciences, University of Birmingham, Birmingham, UK.
2. Oncology, Birmingham Women's and Children's NHS foundation trust, Birmingham, United Kingdom.
3. RRPPS, University Hospitals Birmingham NHS foundation trust, Birmingham, United Kingdom.
4. Radiology, Birmingham Women's and Children's NHS foundation trust, Birmingham, United Kingdom.
5. Department of Psychology, School of Life and Health sciences, Aston University, Birmingham, United Kingdom.
6. Radiology, Alder Hey Children's NHS foundation trust, Liverpool, United Kingdom.
7. Institute of Translational Medicine, University of Liverpool, Liverpool, United Kingdom.
8. The Children's Brain Tumour Research Centre, University of Nottingham, Nottingham, United Kingdom.
9. Sir James Spence Institute of Child Health, Royal Victoria Infirmary, Newcastle upon Tyne, United Kingdom
10. Neuroradiology, Royal Victoria Infirmary, Newcastle Upon Tyne, United Kingdom.
11. Institute of Digital Healthcare, WMG, University of Warwick, Coventry, United Kingdom.
12. Sir Peter Mansfield Imaging Centre, University of Nottingham Biomedical Research Centre, Nottingham, United Kingdom.
13. NIHR Nottingham Biomedical Research Centre, Nottingham, United Kingdom.

**Acknowledgements**

We would like to acknowledge funding from the Cancer Research UK and EPSRC Cancer Imaging Programme at the Children's Cancer and Leukaemia Group (CCLG) in association with the MRC and Department of Health (England) (C7809/A10342), the Cancer Research UK and NIHR Experimental Cancer Medicine Centre Paediatric Network (C8232/A25261), the Medical Research Council – Health Data Research UK Substantive Site and Help Harry Help Others charity. Professor Peet is funded through an NIHR Research Professorship, NIHR-RP-R2-12-019. Stephen Powell gratefully acknowledges financial support from EPSRC through a studentship from the Physical Sciences for Health Centre for Doctoral Training (EP/L016346/1). Professor Theodoros N Arvanitis is partially supported by Health Data Research UK, which is funded by the UK Medical Research Council, Engineering and Physical Sciences Research Council, Economic and Social Research Council, Department of Health and Social Care (England), Chief Scientist Office of the Scottish Government Health and Social Care Directorates, Health and Social Care Research and Development Division (Welsh Government), Public Health Agency (Northern Ireland), British Heart Foundation and Wellcome Trust. We would also like to acknowledge the MR radiographers at Birmingham Children's Hospital, Alder Hey Children's Hospital, the Royal Victoria Infirmary in Newcastle and Nottingham Children's Hospital for scanning the patients in this study. We would also like to thank Selene Rowe at Nottingham University Hospitals NHS Trust for help with gaining MRI protocol information. Dr James Grist is funded by the Little Princess Trust (CCLGA 2017 15).

**Corresponding author:**

Professor Andrew C. Peet

Children's Brain Tumour Research Team, 4th Floor Institute of Child Health

Birmingham Women's and Children's Hospital NHS Foundation Trust

Steelhouse Lane

Birmingham B4 6NH

Email: a.peet@bham.ac.uk

Tel: +44 (0) 121 333 8412, Fax: +44 (0) 121 333 8701

**Abstract**

The imaging and subsequent accurate diagnosis of paediatric brain tumours presents a radiological challenge, with magnetic resonance imaging playing a key role in providing tumour specific imaging information. Diffusion weighted and perfusion imaging are commonly used to aid the non-invasive diagnosis of children's brain tumours, but are usually evaluated by expert qualitative review. Quantitative studies are mainly single centre and single modality.

The aim of this work was to combine multi-centre diffusion and perfusion imaging, with machine learning, to develop machine learning based classifiers to discriminate between three common paediatric tumour types.

The results show that diffusion and perfusion weighted imaging of both the tumour and whole brain provide significant features which differ between tumour types, and that combining these features gives the optimal machine learning classifier with >80% predictive precision. This work represents a step forward to aid in the non-invasive diagnosis of paediatric brain tumours, using advanced clinical imaging.

**Introduction**

Brain tumours are the most common solid tumours in children, accounting for approximately 25% of all childhood cancers. Magnetic resonance imaging (MRI) is commonly performed for children suspected of having a brain tumour at presentation. Challenges are faced by paediatric radiologists to diagnose paediatric brain tumour type using MRI, especially in tumours which do not enhance with gadolinium contrast agent (a significant fraction in paediatric radiology)(Koob and Girard, 2014). Therefore, if a combination of imaging methods can be used to quantify tumour cellular microstructure and perfusion, it may be possible to discriminate between low and high grade, as well as key tumour subtypes such as Pilocytic Astrocytoma, Ependymoma, and Medulloblastoma. Magnetic resonance spectroscopic methods have been shown to be highly predictive in discriminating between tumour types, however this technique is challenging to acquire in regions of the brain with poor magnetic field homogeneity and small lesions(Faghihi et al., 2017; Lin and Chung, 2014). Therefore, other more commonly used imaging-based methods, such as diffusion and perfusion imaging, may be favourable to discriminate between tumour types in the paediatric brain.

Diffusion weighted imaging (DWI) and dynamic susceptibility contrast imaging (DSC) are two advanced magnetic resonance imaging (MRI) techniques available to understand tissue microstructure and perfusion on a cellular and tissue level(Goo and Ra, 2017; Shah et al., 2016; Zhou et al., 2011). These techniques have been used extensively to understand the role of cellularity and microvascular perfusion, in both paediatric and adult brain tumours(Hales et al., 2019; Poussaint et al., 2016), with strong correlations with histology for the aforementioned. DWI utilises diffusion sensitising preparation gradients to remove signal from static water compartments in the brain, producing images weighted by the speed of water motion in a given voxel. With the assumption of Brownian motion, diffusion weighted images can be used to calculate an 'apparent diffusion coefficient (ADC) map', with each voxel value corresponding to the ADC in the voxel ($mm^2s^{-1}$)(Le Bihan, 2003).

DSC is used to spatially image the dynamics of a gadolinium containing contrast agent, using fast imaging techniques such as echo-planar imaging (EPI) and (PRESTO). Data are processed using non-linear fitting techniques to extract uncorrected cerebral blood volume (UCBV), leakage coefficient (K2) and corrected cerebral blood volume (CCBV) maps. CBV maps can then be analysed to quantify the perfusion in a given region of the brain(Shiroishi et al., 2015). DSC has shown to be useful in quantifying perfusion differences between low- and high-grade tumours, as well as in stroke(Boada et al., 2005; Saenger and Christenson, 2010; Sanak et al., 2009).

Supervised machine learning utilises data features (for example mean ADC or mean CBV) and classes (for example 'high and low grade' or tumour types) and to train mathematical algorithms (commonly based on linear algebra) to automatically assign data sets to classes. The ability of a learning algorithm to discriminate between classes can be quantitatively determined using methods such as 'cross-validation'(Erickson et al., 2017). Previous results have shown the ability of supervised methods to separate between tumour subtypes and high/low grade tumours using magnetic resonance spectroscopy, with 1.5T and 3T results showing 79% and 86% balanced accuracy rate (BAR), respectively (Vicente et al., 2013; Zarinabad et al., 2018).

Applications of supervised learning to oncological medical imaging have commonly utilised single measures of the tumour microenvironment (such as image texture, ADC, perfusion, or spectroscopy) to discriminate between tumour types(Fetit et al., 2018; Gill et al., 2014; Orphanidou-Vlachou et al., 2014, 2013; Zarinabad et al., 2017). However, in this study, we hypothesise that combining ADC and perfusion data from tumour Region of Interest (ROI) and the whole brain, provides an increased accuracy for discriminating between low- and high-grade tumours, as well as between tumour sub-types, in comparison to ROI or whole brain measures alone.

## Methods

**Patient recruitment**

49 participants with suspected brain tumours (medulloblastoma (N = 17), pilocytic astrocytoma (N = 22), ependymoma (N = 10)) were recruited from 4 clinical sites in the United Kingdom (Ethics reference: 04/MRE0/41, Birmingham Children's Hospital, Newcastle Royal Victoria Infirmary, Queen's Medical Centre, Liverpool Alder Hey). Participants underwent MRI, discussed below, before invasive biopsy to confirm diagnosis.
All Ependymoma and Medulloblastoma cases were considered high grade, and Pilocytic Astrocytoma as low grade.

**Magnetic resonance imaging**

The imaging protocol for all participants was performed either at 3 or 1.5T and included standard anatomical imaging ($T_1$-weighted, $T_2$-weighted, $T_2$-FLAIR, $T_1$-post contrast), as well as diffusion weighted and dynamic susceptibility contrast, covering the tumour volume (imaging sequence and cohort details found in supplementary Table 1).

**Image post-processing and analysis**

Apparent diffusion coefficient maps were calculated from diffusion weighted imaging, using a linear fit between the two b-value images. DSC time-course data were processed using a gamma-variate fit to form UCBV maps. A leakage correction was undertaken to produce CCBV and K2 maps(Shiroishi et al., 2015). The root mean squared error of the gamma variate fit was used to mask noise and masking any absolute CBV value greater than 3.0 mL $100g^{-1}$ $min^{-1}$. Brain masking, including removal of background and the skull, was performed during the fitting process. CBV maps were normalised to normal appearing white matter. $T_2$- weighted, ADC, and $T_1$-post contrast images were registered to the first DSC volume with SPM12 (UCL), and tumour regions of interest drawn on $T_2$ weighted imaging.

Image analysis, performed in Matlab (2018b, The Mathworks, MA), consisted of calculating the image mean, standard deviation, skewness, and kurtosis on a volume by volume basis for diffusion and perfusion imaging maps maps for regions of interest and the whole brain. Tumour volume ($cm^3$) was calculated from the $T_2$ ROI masks.

**Statistical analysis**

Imaging features were tested for normality using a Shapiro-Wilk test in R (3.6.1) with subsequent ANOVA/Kruskal-Wallace and Tukey post-hoc tests performed to assess for differences in imaging features between low- and high-grade groups, and between tumour types. Receiver Operator Curves (ROC) were defined from significant imaging components for comparison of low versus high-grade tumours, and the area under the curve (AUC) calculated. Statistical significance was determined at $p < 0.05$, with Bonferroni correction for multiple comparisons.

**Machine learning**

The discriminant ability of classifiers, described below, was assessed using the F-statistic (a measure of sensitivity and specificity of the learner) and the between group average precision (the average precision of the learner to correctly classify tumour types), after stratified 3-fold cross validation. Individual tumour group accuracy and F-statistics were also calculated. A flowchart demonstrating the processing pipeline is found in Figure 1.

Tumour volume, ADC and DSC region of interest and whole brain features were processed using principal component analysis to reduce dimensionality, aiming for 95% data variance or N-1 components where not possible (where N is the size of the smallest group). Supervised machine learning was performed using the Orange toolbox (Orange) in Python (3.6), using a Neural Network, AdaBoost, random forest, a support vector machine, and k nearest-neighbours.
Learning algorithms were initialised to first discriminate between low- and high-grade groups, and then between tumour sub-types. The Balanced Accuracy Rate (BAR), F-statistic, and individual group accuracies were calculated for each learner after stratified cross-validation.
A further approach to dimensionality reduction was undertaken by independently performing univariate statistical analysis (described above in 'statistical analysis' section) on 3 stratified subsets of the imaging data (75:25% training:test set size). Features with the highest AUC over all subsets were selected for learning. This combination of features is termed here as the 'univariate' classifier.

**Data oversampling**

Two oversampling methods: data replication and SMOTE (Chawla et al., 2002; Zarinabad et al., 2018), were used to increase the ependymoma group size of training sets by 100%. The oversampled data was processed, with supervised learning, as above and results compared with no oversampling.

## Results

Example DSC and DWI imaging is shown in Figure 2: $T_2$-weighted (A), ADC(B), Uncorrected CBV (C), K2 (D), and Corrected CBV (E).

**Tumour Region of interest and whole brain analysis reveals features which differ between low- and high-grade tumours and some tumour types**

Region of interest and whole brain features analysis revealed a number of imaging features that were significantly different between low- and high-grade tumours. With ADC mean having the highest AUC of 0.8, with a range of 0.37 to 0.78 AUC for other features.

Further to distinguishing between low- and high-grade tumours, significant differences in ADC features were observed between Pilocytic Astrocytomas and Medulloblastomas: ADC ROI mean ($1.5 \pm 0.3$ vs $0.9 \pm 0.2$ mm$^2$ s$^{-1}$, p < 0.001, ACU = 0.75), ADC ROI skewness ($0.9 \pm 1.0$ vs $1.9 \pm 0.9$, p = 0.006), and ADC ROI kurtosis ($5 \pm 3$ vs $9 \pm 5$, p = 0.045, ACU = 0.65). A significant difference in tumour volume between Pilocytic Astrocytomas and Ependymomas was observed ($2.3 \pm 3.1$ vs $9.0 \pm 11.2$ cm$^3$, respectively, p = 0.02, AUC = 0.67).

Whole brain analysis revealed a significant difference between high and low grade tumours; ADC mean ($0.68 \pm 0.24$ vs $0.9 \pm 0.2$ mm$^2$ s$^{-1}$, p = 0.001, AUC = 0.77), and uncorrected CBV whole brain mean ($0.11 \pm 0.03$ vs $0.13 \pm 0.02$ mL 100g$^{-1}$ min$^{-1}$, p = 0.002, AUC = 0.62). Pilocytic Astrocytomas and Medulloblastomas also differed in the whole brain features such as corrected CBV mean ($1.1 \pm 0.3$ vs $1.2 \pm 0.2$ mL 100g$^{-1}$ min$^{-1}$, respectively, p = 0.009, AUC = 0.62 and ADC mean ($0.9 \pm 0.2$ vs $0.7 \pm 0.3$ mm$^2$ s$^{-1}$, respectively, p < 0.001, AUC = 0.78). Full tumour subtype results are shown in Table 1.

**Supervised learning can distinguish between low- and high-grade tumours and different tumour types with a combination of region of interest and whole brain features.**

To discriminate between tumour types, the univariate classifier performed the best (features: ADC ROI mean, ADC whole brain kurtosis, uncorrected CBV ROI skewness, and tumour volume) using an AdaBoost learner (precision = 86%, F-statistic = 0.85). Excluding ADC whole brain kurtosis from the above classifier resulted in a reduction to 82% precision F-statistic = 0.75.

Utilising PCA to reduce dimensionality did not perform as well as the univariate classifier, with BAR ranging from 66%-64% (all imaging features and all ROI features, respectively). All results including individual class precision are detailed in Tables 2A and B.

A combination of all ROI features had the highest precision to discriminate between high- and low-grade tumours with a support vector machine (86% precision, 11 principal components). All other results for high-low grade classification are shown in supplementary table 3.

**Oversampling increases learner accuracy for some classifiers**

Oversampling increased classifier precision for PCA based classifiers, as demonstrated by large increase in BAR, results shown in Table 3A, however, it did not increase BAR for the univariate classifier (85% vs 85% vs 86%, no oversampling vs data replication vs SMOTE, respectively). Indeed, using oversampling methods with the univariate classifier showed an increase in the classification accuracy for Ependymomas, but little change for medulloblatomas and a decreased accuracy for pilocytic astrocytomas. All group precision and F-statistic results are presented in Table 3B.

**Discussion**

This study has demonstrated that a combination of multiparametric MRI, univariate analysis, and machine learning techniques can be employed to distinguish between both high- and low-grade paediatric brain tumours, as well as enabling tumour type classification with high accuracy (achieving 86% BAR).

Previous studies have commonly focused on the use of a single data type (for example ADC or DSC perfusion measures) to discriminate between high- and low-grade tumour types as well as common tumour histological diagnoses, with results in this study agreeing with previous findings(Bull et al., 2012).

Studies of diagnostic classifiers for tumours based on imaging have concentrated on using data from regions of interest drawn around the tumour or regions of abnormality on the conventional MRI. Here we have also investigated imaging features selected from the whole brain and shown that these are significantly different between the tumours. In addition we found that these features can improve the accuracy of the diagnostic classifier when included in it.

A particularly interesting result of this study showed that feature selection, informed by univariate statistics, provided a classifier that outperformed other methods. This emphasises the importance of optimising the way in which features are selected for input to the machine learning classifier. Large numbers of features cannot be used due to the risk of over-fitting and, consequently, obtaining over optimistic estimates of the accuracy of the classifier. However, methods, such as principal component analysis, which select features by how much variability exists in the data, may not select the most discriminatory features but those that vary most throughout the data set used in this study.

Previous MR spectroscopic studies at 1.5 and 3T utilising supervised machine learning have achieved similar results as demonstrated here(Vicente et al., 2013; Zarinabad et al., 2017) and it would be interesting to determine the added value of combining these modalities. Challenges are faced in the acquisition of DSC data, particularly the injection of gadolinium in a highly regulated manner in children and the use of arterial spin labelling. These may present an alternative option for utilising perfusion imaging in the future(Novak et al., 2019; Radbruch et al., 2015).

Challenges are faced in paediatric oncological studies with low recruitment rates, due to the low disease incidence in the population. Therefore, multi-centre approaches present an opportunity to both collect the data sets required to undertake machine learning approaches, as well as increasing the statistical power of the study itself. Here we have also shown that multi-parametric data from multiple centres can be combined to form powerful classifiers in the study of paediatric brain tumours.

Work beyond this study could focus on the expansion to other less common brain tumour sub-types, such as such as genetic subtypes of Medulloblastomas, to extend the relevance and scope of this work. This, in turn, will aid in the radiological classification and diagnosis of many other tumour types beyond the main three represented in this study. Furthermore, the addition of other microstructure data, such as diffusion kurtosis and intra-voxel incoherent motion, may provide further information regarding the tumour microenvironment, and, therefore, further aid in the discrimination between tumour types.

Limitations of this study include low participant numbers in the Ependymoma group, a common challenge in paediatric imaging studies. This was mitigated, to some extent, by the use of oversampling in the machine learning classifiers, although further numbers in this group should be obtained. Overall, classifier results have shown the power of machine learning to distinguish between tumour types.

In conclusion, this study has demonstrated the power of combining advanced MRI methods with machine learning to provide a non-invasive diagnosis of paediatric tumour types.

**Figure captions**

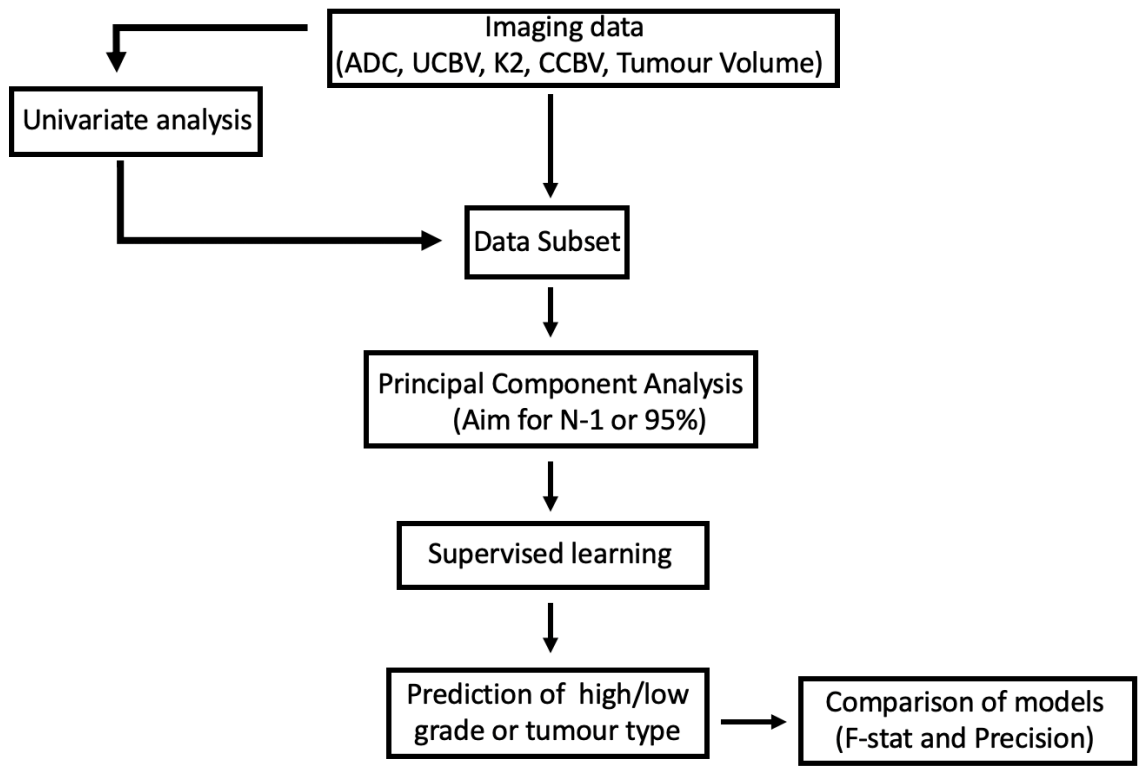

**Figure 1 – Data processing pipeline used in this study.**

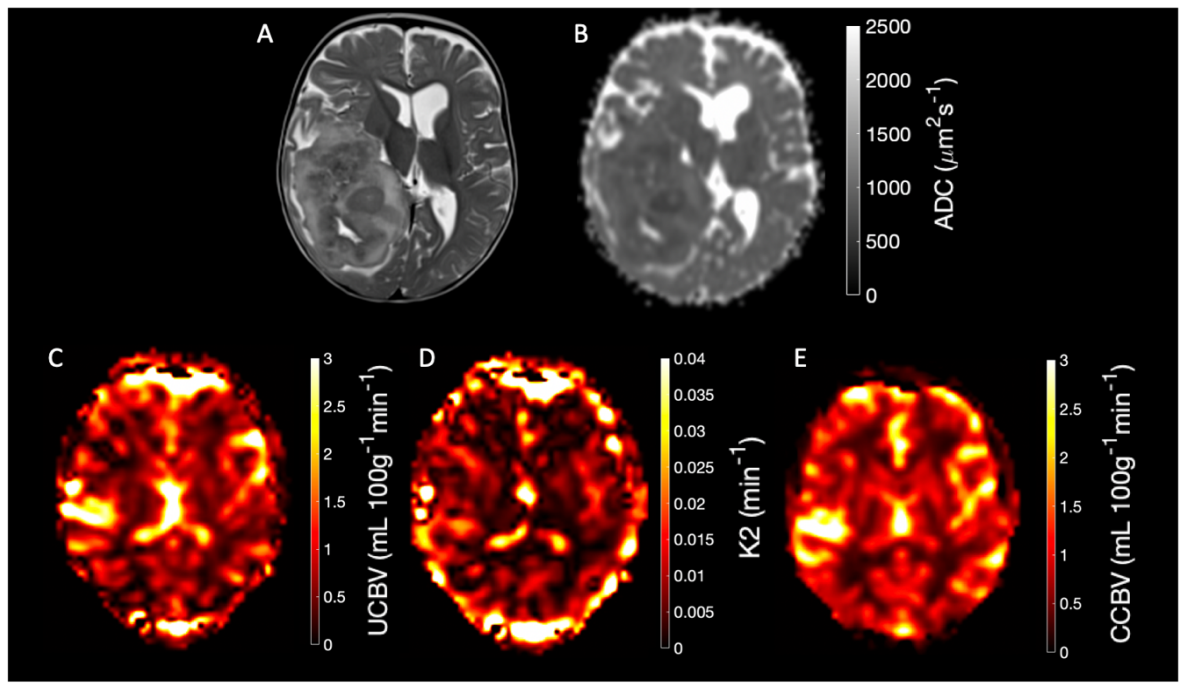

**Figure 2 – Example anatomical, perfusion and diffusion maps of an Ependymoma.** A) T2-weighted imaging, B) Apparent Diffusion Coefficient map, C) Uncorrected Cerebral Blood Volume map, D) K2 map, E) Corrected Cerebral Blood Volume map.

| Feature | Pilocytic Astrocytoma | Ependymoma | Medulloblastoma |
|---|---|---|---|
| ADC ROI Mean ($mm^2s^{-1}$) | 1.5 ± 0.4* | 1.2 ± 0.1 | 0.9 ± 0.2 |
| ADC ROI Skewness | 0.9 ± 1.0* | 2.0 ± 1.0 | 1.9 ± 0.9 |
| ADC ROI Kurtosis | 5 ± 3* | 8 ± 5 | 9 ± 5 |
| ADC WB Mean ($mm^2s^{-1}$) | 0.9 ± 0.2* | 0.7 ± 0.3 | 0.6 ± 0.2 |
| ADC WB Skewness | 1.2 ± 0.3* | 1.5 ± 0.5 | 1.6 ± 0.5 |
| ADC WB Kurtosis | 5 ± 1* | 6 ± 2 | 7 ± 2 |
| CCBV WB Mean ($mL\ 100g^{-1}\ min^{-1}$) | 1.1 ± 0.3* | 1.2 ± 0.2 | 1.3 ± 0.2 |
| Tumour volume ($cm^3$) | 2.3 ± 3.1 ** | 9.0 ± 11.2 | 3.3 ± 2.3 |

**Table 1 – Univariate tumour separation results.** A number of imaging features were found to be significant (* = Pilocytic Astrocytoma vs Medulloblastoma at $p < 0.05$, ** = Pilocytic Astrocytoma vs Ependymoma at $p < 0.05$).

A

| Average Learner | All features (9 PCs) | ROI features (9 PCs) | Whole brain features (9 PCs) | Univariate ROI | Univariate all features |
|---|---|---|---|---|---|
| AdaBoost | 62%, 0.61 | 70%, 0.67 | 66%, 0.62 | 78%, 0.76 | 86%, 0.84 |
| Random Forest | 64%, 0.64 | 71%, 0.72 | 54%, 0.55 | 75%, 0.72 | 77%, 0.73 |
| Support Vector Machine | 62%, 0.60 | 67%, 0.73 | 54%, 0.55 | 62%, 0.67 | 79%, 0.75 |
| K Nearest Neighbors | 66%, 0.58 | 55%, 0.55 | 50%, 0.46 | 82%, 0.75 | 82%, 0.72 |
| Neural Network | 56%, 0.55 | 68%, 0.66 | 50%, 0.49 | 61%, 0.64 | 77%, 0.74 |

B

| Class | All features (kNN) | ROI features (Random Forest) | Whole Brain features (AdaBoost) | Univariate (AdaBoost) |
|---|---|---|---|---|
| Pilocytic Astrocytoma | 65%, 0.64 | 76%, 0.78 | 77%, 0.74 | 95%, 0.86 |
| Medulloblastoma | 50%, 0.59 | 75%, 0.81 | 50%, 0.59 | 74%, 0.85 |
| Ependymoma | 100%, 0.4 | 50%, 0.33 | 67%, 0.36 | 83%, 0.71 |

**Table 2 - Supervised learning results for tumour type discrimination.**

A

| Sampling method | All features (15 PCs) | ROI features (11 PCs) | Whole brain features (9 PCs) | Univariate |
|---|---|---|---|---|
| Normal | 66%, 0.58 kNN | 71%, 0.72 RF | 66%, 0.62 AdaBoost | 85%, 0.84 AdaBoost |
| Data replication | 82%, 0.87 RF | 78%, 0.78 AdaBoost | 80%, 0.76 AdaBoost | 85%, 0.85 RF |
| SMOTE | 85%, 0.84 AdaBoost | 82%, 0.80 AdaBoost | 78%, 0.78 AdaBoost | 86%, 0.85 AdaBoost |

B

| Class | Univariate with no oversampling | Univariate with data replication | Univariate with SMOTE oversampling |
|---|---|---|---|
| Pilocytic Astrocytoma | 95%, 0.86 | 91%, 0.91 | 86%, 0.82 |
| Medulloblastoma | 74%, 0.85 | 75%, 0.75 | 87%, 0.84 |
| Ependymoma | 83%, 0.71 | 85%, 0.85 | 83%, 0.88 |

**Table 3 – Results containing standard, random, and SMOTE oversampling of ependymoma features.** Balanced accuracy rate for all, ROI, whole brain, and optimized features shown in A, and group classification results from best performing classifier in B (precision (%), F-statistic).

| | Tumour type | Male:Female | Imaging Field strength |
|---|---|---|---|
| A | Pilocytic Astrocytoma | 9:13 | 1.5/3T |
| | Medulloblastoma | 9:8 | 1.5/3T |
| | Ependymoma | 5:5 | 1.5/3T |

| | Imaging sequence | Voxel volume (mm$^3$) | Flip angle (degrees) | Repetition time (ms) | Echo time (ms) | B-values | Temporal resolution (s) |
|---|---|---|---|---|---|---|---|
| B | T$_2$-weighted | 1.6-2.6 | 90 | 3000-5230 | 80-108 | N/A | N/A |
| | Diffusion weighted imaging (DWI) | 25-45 | 90 | 2600-5000 | 48-68 | 0, 800 or 1000 | N/A |
| | DSC imaging (Echo planar imaging) | 11-45 | 20-75 | 582-2343 | 18.4-40 | N/A | 0.6-2.3 |
| | DSC imaging (sPRESTO) | 18-22 | 7 | 15.5-17.2 | 23.5-25.2 | N/A | 1.2-1.6 |

Supplementary table 1 – Cohort (A) and imaging parameters (B) used in this study.

| Imaging feature | Low-grade | High-grade | AUC |
|---|---|---|---|
| UCBV ROI Mean (mL 100g$^{-1}$ min$^{-1}$) | 0.4 ± 0.8 | 0.9 ± 0.9 | 0.68 |
| UCBV ROI Skewness | 0.0 ± 0.8 | -0.5 ± 0.9 | 0.69 |
| ADC ROI Mean (mm$^2$s$^{-1}$) | 1.46 ± 0.40 | 0.10 ± 0.23 | 0.80 |
| ADC ROI Kurtosis | 5 ± 3 | 9 ± 5 | 0.75 |
| ADC ROI Skewness | 0.8 ± 1 | 1.9 ± 1.0 | 0.78 |
| WB ADC Mean | 0.68 ± 0.24 | 0.90 ± 0.20 | 0.77 |
| WB ADC Kurtosis | 5 ± 2 | 7 ± 2 | 0.77 |
| WB ADC Skewness | 1.2 ± 0.3 | 1.5 ± 0.6 | 0.37 |
| WB UCBV Mean (mL 100g$^{-1}$ min$^{-1}$) | 0.11 ± 0.03 | 0.13 ± 0.02 | 0.62 |
| WB UCBV STD (mL 100g$^{-1}$ min$^{-1}$) | 1.3 ± 0.4 | 1.0 ± 0.4 | 0.60 |

**Supplementary Table 2 – Significant univariate results from high/low grade separation.** Analysis showed a number of significant ADC and DSC imaging features between low and high grade groups. AUC = Area under the curve.

| Feature combination | Number of Principal Components | Optimal learner | Precision (%) | F-statistic |
|---|---|---|---|---|
| Univariate all features | N/A | AdaBoost | 83 | 0.84 |
| All features | 17 | Random Forest | 82 | 0.82 |
| All ROI features | 11 | Support Vector Machine | 86 | 0.86 |
| All Whole Brain features | 9 | Support Vector Machine | 74 | 0.74 |

**Supplementary table 3 – Supervised learning results for low/high grade.** Results showed that a PCA reduced combination of ROI features combined with a Support Vector Machine provided the best learner to discriminate between high and low grade tumours.